\documentstyle[epsfig,aps]{revtex}

\begin{document}

\draft

\title{\bf Various Spin Phases  from
 Lightly-Doped Two-legged $t$-$J$ Spin Ladders}
\author{Yujoung Bai}
\address{Center for  Strongly-Correlated Materials Research, 27-113, Seoul National University, Seoul 151-742, Korea}
 \author{Sung-Ik Lee}
\address{National Creative Research Initiative Center for Superconductivity, 
Department of Physics,
Pohang University of Science and Technology, Pohang, Kyungbuk  790-784, Korea}

\maketitle
\begin{abstract}

We   study   two-leg   spin  ladder   systems analytically by using a $t$-$J$  model 
 which includes the interchain spin exchange and the interchain hopping term. 
  The spin part is mapped to  
 a quantum sine-Gordon model via a bosonization method that fully accounts for
 the phase fluctuation. The spin gap in the Luther-Emery phase evolves to
 gapless Luttinger liquid phases in even-ladders at certain doping concentration.
  This transition occurs as doping rate increases and/or
  the ratio of the interchain spin exchange and hopping to the
  intrachain couplings decreases. We also estimate
the transition temperature at which the conventional
electron phase can be  deconfined to spinon and holon. 
\\
\end{abstract} 


\section{ introduction}

Low-dimensional quantum magnetic systems, such as spin chains, spin ladders, and planar
lattices provide very useful information about the physics of high $T_C$ superconductors.
During the past few years, spin-ladder systems have been studied extensively by various 
groups\cite{capponi}\cite{hlubina}\cite{shelton}, especially by numerical methods or by mean-field (MF) calculations\cite{dagotto}\cite{white}.
A``spin ladder'' consists of  parallel ``chains'' of magnetically bound $Cu$ atoms
 connected by interchain
couplings (``rungs''), as  found in $Sr_{n-1} Cu_{n+1} O_{2n}$,
$La_{4+4n}Cu_{8+2n}O_{14+8n}$ (Fig. 1).
 Spin ladder compounds may become
superconductors if doped with charge carriers\cite{kumagai}\cite{takagi}.
 Recently, superconductivity has been found in
$Sr_{14-x}Ca_{x}Cu_{24}O_{41}$\cite{hiroi} 
 below 12 K and with external pressure of 3 GPa.

 Beginning with the half-filled ground state of a spin liquid, short-range singlets
  are formed predominantly across the rungs 
 of the ladders  from the antiferromagnetic (AF) states. 
  The unusual spin-liquid nature of an undoped parent system and the evolution 
 of the finite gap upon doping have also been studied by numerical methods
 \cite{hayward}. Excitations of the singlet ground state yield
  spin triplets  which propagate along the ladder direction. The study
  of even-leg ladder systems has shown 
  that spin gaps decrease as the number of legs increase.
   ( The gap remains finite down to 
  $J_{\perp}= J_{\parallel}$, which is a signature of 
 superconductivity\cite{troyer}). Within the mean-field theory, in numerical works and
  renormalization group (RG) calculations, a superconducting phase with a d-wave nature
   emerges from the short-range RVB liquid upon doping.
    Then, new quasiparticle (QP) 
  states appear which carry both charge (positive) and spin. 
  
  In the weak-coupling approach, one starts with two non-interacting spin chains and
  turns on 
 perturbative interchain interactions. For finite interchain interactions, this
 method does not  accurately describe the correlations near the phase transitions. 
   Zhang et al.\cite{Zhang} have studied two-leg $t-J$ ladders {\it without} 
   the interchain exchange by mapping the spin part to a sine-Gordon model.
   Our study includes the spin exchange interaction across
 the rungs of the spin ladders and is based on a (nearly) exactly solvable model.
 Based on the soliton description, we discuss the phases and possible
  transitions between them. Exactly solving for spin ladders  gives better
  understanding of the correlations between phase excitations and topological 
  defects, while comparing with some weak-coupling studies\cite{fabrizio} and  MF results which partly addressed this issue\cite{sigrist}.

 	We use a $t$-$J$ model for two-leg ladders in the strongly(isotropically) coupled
 regime in which the interchain interactions ($t_\perp$, $J_{\perp}$) are larger(equal)
 than the intrachain interactions $(t_{\parallel}, J_{\parallel})$. Introducing weak 
 coupling between two-leg {\it ladders} 
 is staightforward for even-leg ladder systems. Recent studies\cite{rice} have
 shown that ladder systems with an odd number of legs are  
conducting Luttinger liquids  in the odd-parity channel and 
  spin liquids in the even parity channel\cite{sigrist}. 
  Odd-legged ladders can be mapped to a single-chain system while even-legged
   ladders are equivalent to a double-chain ladder.
   
   \section{ the model}
   
 An effective model for the $t$-$J$ Hamiltonian is obtained as follows :
\begin{eqnarray}
H & = & -t_{\parallel}\sum_{l=1,2}\sum_{<ij>,\sigma}(
C_{li\sigma}^{+}C_{lj\sigma} + h.c.)\nonumber \\
& - & t_{\perp}\sum_{l,l^{\prime}=1,2}\sum_{<ij>,\sigma}(
C_{li\sigma}^{+}C_{l^{\prime}i\sigma} + h.c.)\nonumber \\
 & + & J_{\parallel}\sum_{l=1,2}\sum_{<ij>,\sigma}( S_{li\sigma}S_{lj\sigma}
-{1\over 4} n_{li\sigma}n_{lj\sigma})\nonumber \\
& + & J_{\perp}\sum_{l,l^{\prime}=1,2}\sum_{<ij>,\sigma}(S_{li\sigma}S_{l^{\prime}i\sigma}
 -  {1\over 4}n_{li\sigma}n_{l^{\prime}i\sigma})\nonumber \\
& + & \mu\sum_{l=1,2}\sum_{\sigma} C_{li\sigma}^{+}C_{li\sigma}\end{eqnarray}
where $<ij>$ is for nearest-neighbor( $j=i+1$) interactions, $l$ is the leg index
 and $\sigma$ is the spin $(\uparrow \downarrow)$ index.
 In  strongly coupled $t$-$J$ models with a large Coulomb repulsion (large
 $U/t$ and large
$J_{\perp}/J_{\parallel})$, we impose the single-occupancy constraint 
$\sum_{\sigma}C_{i\sigma}^{+}C_{i\sigma}=0 \hspace{0.3cm}or\hspace{0.3cm} 1$
at each site.
The electron chemical potential $\mu$ is added for the constraint in the form of 
 a Lagrange multiplier.  The charge(hole) concentration is related to
$\mu (1-\delta)= {1/({1 + e^{-\beta\mu}}})$.
In order to study the spin part and the charge part separately, we write 
 the electron operator as the product of a charge carrying holon $h_{i}$ and
  a pseudospin operator  $S_{i}$:
$C_{i\downarrow}=Ph_{i}S_{i}^{+}P^{+}\equiv h_{i}S_{i}^{+}, \hspace{0.1cm}
C_{i\uparrow}=Ph_{i}S_{i}^{-}P^{+}\equiv h_{i}S_{i}^{-}.$
 The projection operator $P= (1-n_{i,-\sigma})$ reduces the four on-site states 
 $|0,\uparrow>, |0,\downarrow>, |1,\uparrow>$ and $|1,\downarrow>$ to the
  physical Hilbert spaces $|hole>, |\uparrow>$ and $|\downarrow>$, thereby 
  imposing  single-occupancy at each site. Since the charge sum rule
 $(\sum_{\sigma} C_{i\sigma}^{+}C_{i\sigma}=1-\delta)$ holds when
 the projection operators are neglected,
  we equate the electron operator with $ h\bigotimes S $. 
 The pseudospin operators acting at site $i$ turn into  spinless
  fermion operators
  $a$ via a Jordan-Wigner
    transformation.
	
\section{ continuum limit and bosonization}
	
	 In the spin exchange part, the pseudospin operators anticommute on the same site and commute on the different sites. so, we have $H(J)$ parts become
	 $J_{\parallel}\sum_{i=1}^N(S_{i}^{+} S_{i+1}^{-} + h.c.)$ and
	    $J_{\perp}\sum_{i,l}(S_{i,l}^{+} S_{i,l+1}^{-} + h.c.)$.
	  From Jordan-Wigner transformation, the intrachain term $S_{i}^{+} S_{i+1}^{-}=a_i^+\exp[-i\pi a_i^+a_i]a_{i+1}=
	  a_i^+(1-2a_i^+a_i)a_{i+1}$ for spin $1\over 2$-system, while
	  the interchain term 
	   $S_{i,l}^{+} S_{i,l+1}^{-}=a_{i,l}^+a_{, i,l+1}-2a_{i,l}^+a_{i,l}^+a_{i,l}a_{i,l+1}$.
	   We introduce spinor fields $\phi_\alpha$ for the slowly-varying field $\widetilde{a_i}=a_i/i^n$ where $\alpha$=even at even(2s) site $i$ and odd at odd(2s+1) site $i$. The even-site and odd site are coupled by nearest-neighbor interaction, yielding $\sum_{i=1}^N \widetilde{a}_{i}^+\widetilde{a}_{i+1}=
	   \sum_{s=0}^{N\over 2}[ \phi_e^+(2s)\{\phi_o(2s+1)-\phi_o(2s-1)\}+
	   \phi_o^+(2s+1)\{\phi_e(2s)-\phi_e(2s-2)\}]$. In continuum, the  fermion field is given $\phi_\alpha(s)=\sqrt{2a_0}\Psi_{\alpha}(x)$ where $x=2sa_0$ with the lattice constant $a_0$. Then, $\phi_o(2s+1)-\phi_o(2s-1)\approx 2a_0\sqrt{2a_0}\partial_x\Psi_0(x)$. The commutation relations are 
	    $\{\Psi_{\alpha}(x), \Psi_{\beta}(y)\}=\delta_{\alpha\beta}\delta(x-y)$ and 
	   $\{\phi_{\alpha}(n), \phi_{\beta}(n\prime)\}=\delta_{\alpha\beta}\delta_{nn\prime}$.
	    With the relation for exponentiated operators
    $e^{A}e^{B}=e^{[A+B]}\hspace{0.1cm}e^{-[A,B]/2}
  =e^{B}\hspace{0.1cm}e^{A}\hspace{0.1cm}e^{-[A,B]}$ ,
   we  map the spin  exchange interaction terms 
 \begin{eqnarray}
& J_{\parallel} & \sum_{<ij>,\sigma} S_{i\sigma} S_{j\sigma}
 =  J_{\parallel}(S_{i}^{+} S_{j}^{-} + S_{i}^{-} S_{j}^{+}+ S_{iz}S_{jz})\nonumber\\
& = & J_{\parallel}(\widetilde{a}_{i}^{+}\widetilde{a}_{i+1}-2\widetilde{a}_{i}^{+}\widetilde{a}_{i}^{+}\widetilde{a}_{i}\widetilde{a}_{i+1}+ h.c.)
\equiv  H(J_{\parallel}).\end{eqnarray}
  
  Now the intrachain spin exchange
part of Hamiltonian becomes
 \begin{eqnarray}
& H & (J_{\parallel}) =  2 a_{0} J_{\parallel}\int dx \nonumber\\
&[& \Psi^{+}_{\eta}(x) i \sigma^{1} \partial_{x} \Psi_{\eta}(x)
+ \Psi^{+}_{\eta}(x) \Psi_{\eta}(x)\Psi^{+}_{\eta}(x) i \sigma^{1}
\partial_{x} \Psi_{\eta}(x)]\end{eqnarray}
where  $\Psi_{\eta}$ is the Fermi field of two species(bonding, anti-bonding band) and 
$\sigma^{1}$ is the Pauli matrix.

  Then, we  bosonize the Fermi field with phase fluctuation parameters which can
  vary depending on the interactions (rather than the parameter being a constant $\sqrt {4\pi}$). Employing the operator expansion of the Fermi field\cite{fradkin}, 
$\Psi_{\eta}(x)= \exp[i\alpha\phi_\eta(x)]\exp[i\beta\int_{-\infty}^{x}dy
\partial_0\phi_\eta(y_0,y)]$
where the Bose field $\phi$ and its canonical conjugate momentum
 $\Pi=\partial_0\phi$ satisfy  $[\phi(x), \Pi(x^\prime)]=i\delta(x-x^\prime)$ and  
 $[\Pi(x),\partial_0\phi(x^{\prime})]={i\over \pi}\partial_{x}\delta(x-x^{\prime})$.
 In order to satisfy the commutation
relation of $\Psi_{\eta}(x)$, the phase fluctuation parameters $\alpha$ and 
$\beta$ are related as $2\alpha\beta=\pm\pi$.
 The Fourier components of density operators $\rho_\eta(q)=\sum_{k}\Psi_{\eta}^+(k+q)\Psi_{\eta}(k)$ satisfy the boson commutation relation 
$[\rho_R(q), \rho_R(-q\prime)]={-L\over 2\pi}q\hspace{0.1cm}\delta_{qq\prime}$,
$[\rho_L(q), \rho_L(-q\prime)]={L\over 2\pi}q\hspace{0.1cm}\delta_{qq\prime}$ and 
 $[\rho_R(q), \rho_L(-q\prime)]=0$, where 
 $\rho_{R\eta}(x)- \rho_{L\eta}(x)=\Pi_\eta$ and $\rho_{R\eta}(x)+\rho_{L\eta}(x)={-1\over \pi}\partial_x\phi_\eta$ is the fermion density deviation from its average.  Introducing the dual field 
 $\int_{-\infty}^{x}dy \Pi_\eta(y)={1\over \pi}\theta_\eta(x)$ and
 linearizing the four branches of the spectrum near the Fermi surface($\pm k_{F}$), 
  $\Psi_{\eta}(x)=e^{ik_{F}x}R_{\eta}(x)+ e^{-ik_{F}x}L_{\eta}(x)$ with the right-going (R) and left-going (L) components for each specie,
 \begin{eqnarray}
 R_{\eta}(x) & = & {1\over \sqrt{2\pi a_0}}F_{\eta}\hspace{0.1cm}
 \exp[-i(\theta_\eta(x) -\phi_\eta(x))],\nonumber \\
 L_{\eta}(x) & = & {1\over \sqrt{2\pi a_0}}F_{\eta} \hspace{0.1cm}
 \exp[-i(\theta_\eta(x)+\phi_\eta(x))]\end{eqnarray}
 where the Klein factors $F_{\eta}=\exp[-i\theta_\eta]$  are ladder operators\cite{haldane}\cite{von Delft}, raising/lowering the $\eta$-fermion number by one. They are taken to yield the same signs for the species($\eta$) thus ignored in the thermodynamic limit, since the transverse hopping and spin exchange lead to interchain {\it pair} fluctuations\cite{fabrizio} with both specie participating.
Also, the $\Psi$ fields  are
written as even $\Psi_{e\eta}={1\over\sqrt 2}(R_\eta + L_\eta)$ and 
  odd $\Psi_{o\eta}= {1\over\sqrt 2}(-R_\eta + L_\eta)$.
  Resultantly, the intrachain exchange $J_{\parallel}$ maps to the form of 
 $(\partial_{\mu}\phi_\eta)^2 +
 2cos({\beta\over 2}\phi_\eta)$.

 From the  quartic interchain exchange  $J_{\perp}$ part of Hamiltonian,
  $ -2 J_\perp a_{i,l}^{+}a_{i,l}^{+} a_{i,l} a_{i,l+1}=$
  \begin{equation}
   -2i J_\perp (2a_0)^2 [\Psi_{l,e}^+\Psi_{l,e}\Psi_{l,e}^+\partial_{x}\Psi_{l+1,e}
    + \Psi_{l,o}^+\Psi_{l,o}\Psi_{l,o}^+\partial_{x}\Psi_{l+1,o}]
  \end{equation}
   There are actually ten terms giving non-zero contribution to the resultant  sine-Gordon, three of which are shown below.
 \begin{eqnarray}
& H & (J_{\perp}) =  -J_{\perp}\sum_{i,\sigma} S_{1i\sigma}S_{2i\sigma}
=  -2J_{\perp}a_{0}\int dx \nonumber\\
&[& L_{1}^{+}\partial_{x}{R_{2}}
 -R_{1}^{+}{L_{1}}
 R_{1}^{+}\partial_{x}{L_{2}}+
 {R_{2}}^{+}L_{2}L_{2}^{+}\partial_{x}{L_{1}}+..].\end{eqnarray}
 The terms in the integrand above break the left-right (chiral) symmetry 
 which arises from the Umklapp scattering between different chains (and backscattering between opposite spins).
   The interchain hopping term 
  $H(t_\perp) = 
 -t_\perp\sum_{i\sigma}(C_{1i\sigma}^{+}C_{2i\sigma}+H.C.)$ yields identical forms  
  in which we put the {\it  intrachain} charge hopping OP 
  $h_{l,i}^{+}h_{l,i+1}=\alpha\delta\equiv \delta_\parallel$
  where $0 < \alpha < 1$ and {\it interchain} OP
  $h_{l,i}^{+}h_{l+1,i}=\delta[1+ {t_\parallel \over t_\perp}(1-\alpha)]\equiv \delta_\perp$.
   Actual values of $\alpha$, $\delta_{\perp}$, $\delta_\parallel$ and above  rates depend on the interaction strengths $t_\perp$, $t_\parallel$.
   
 Thus, the spin part of the Hamiltonian collected from the intrachain and the
 interchain contributions is 
 \begin{equation}
 H_{s}=
 \int dx\hspace{0.2cm}{1\over {2\pi}^2} 
 [\hspace{0.1cm} (\partial_{\mu}^2{\widetilde\phi_{1}}+
 \partial_{\mu}^2{\widetilde\phi_{2}}) 
 +W cos({\beta\over 2})({\widetilde\phi_{1}}+{\widetilde\phi_{2}})] 
 \end{equation}
 where ${\widetilde\phi}$ is a rescaled field from $\phi=(1+2W\pi)\widetilde\phi$
 with the lattice constant $a_{0}=1$.
 This is a quantum sine-Gordon equation which is the equation of motion of solitons
 (which are  magnons with spin 1 (charge 0) made of 
 two spinons bound by a confining potential).
 
 \section{ phase fluctuation and the interaction strengths}
 
         The constant prefactor is the ratio of interaction strengths
  interchain to intrachain
  \begin{equation}
W  =  {\delta_\perp t_{\perp}-J_{\perp}
\over (\delta_\parallel t_{\parallel}- J_{\parallel})}
\end{equation}
  and is related to the phase fluctuation parameter 
 $\beta^2=16\pi/(1+2W)$.
 In case of frustrated AF (which may occur in the ground state of odd-legged ladders),
 the sine-Gordon equation simplifies to
 $ H={1\over {2\pi}^2}\hspace{0.1cm}[(\partial_{\mu}{\widetilde\phi})^2
 +W\hspace{0.1cm} cos(\beta\widetilde\phi)]$.
 For any finite value of $\beta$, the system has a zero-point fluctuation,
   In the dilute-soliton limit, the
  quantum fluctuation is dominated by the hard-core interaction
   between the spinless fermions\cite{haldane}. 
  Depending on the parameter $\beta$ or $W$,
 we have various soliton phases : 
(1) $W ={J_{\perp}\over J_{\parallel}} \equiv W_{0}$  : 
This is undoped state whose ground state is a AF spin liquid, as known from
 existing numerical studies\cite{troyer}.
(2)$W \ge  W_{0}$  :  Here we have Luther-Emery liquid with the gapped spinon(soliton)
 and gapless charge excitations.  In the lower-temperature regime,
  excitonic soliton-antisoliton bound states exist. These correspond to integer
  spin magnon states  which are bound states of two spin-$1\over 2$ spinons. The CDW
 inherited from chains exists as well as the Luttinger liquid phase holons.
When temperature increases, the excitonic bound states can be broken to spin-triplets
($S=1$ magnon)which make the system spin-gap metal. 
(3) ${1\over 2} < W < W_{0}$ : In this case, 
 repulsive interaction between spinons exists, and there are no stable 
 solitons. This is equivalent to the strong fluctuation of a gauge field. SDW is in the background.
(4) $0 < W \le {1\over 2}$ : The spinons 
  are a Luttinger liquid as shown in algebraically decaying 
  spin-spin correlation given below. The charge part is a Luttinger liquid as in (2).
 (5) $W=0$ or $\delta_{\perp}J_{\perp}= t_{\perp}$ : This limit  corresponds to 
 single soliton(spinon gas) mode.
 (6)$W =\infty $ or $\delta_{\parallel}J_{\parallel}= t_{\parallel}$ : This 
 corresponds to the classical limit with a finite  soliton density. 
  The electron states have dimensional crossovers from the incommensurate 1D regime 
    for $ W < {1\over 2}$ to commensurate 2D-like states for $W_0 < W $. 
	 At low doping  and at low temperatures,   
	  the system can  develop superconducting channel\cite{hayward}. 
	   An excited spin phase generates  magnons from triplets and solitons which propagate in the background of 
	  spin-density-wave(SDW). Weak but finite anisotropy or spontaneous breaking of
	  chiral symmetry cause spin stiffness and low-energy excitations. 
	   The interplay of the excitation modes, such as order-preserving collective magnons and 
	 order-destroying topological solitons, may possibly lead to some 
	 long-range  or intermediate-range order. Considering the asymptotic behaviours shown from scaling equations (Appendix),  a schematic phase diagram is given in Fig. 2, which shows the various spinon 
 and electron phases with the parameter $W$. 
 
 \section{ charge part}
       The separated charge part of the Hamiltonian is diagonalized via a 
Bogoliubov transformation of the holon operators. 
$H  =  \sum_{k}E_{k}(f_{k}^{+}f_{k}+\rm constants)$
 where the holon excitation spectrum 
$E_{k}=-2 t_{\parallel}
cos\hspace{0.1cm}k {+\atop -}
 {\delta_\perp} t_{\perp}$ is valid for
$t_{\perp}\ge t_{\parallel}(1-cos\hspace{0.1cm}2\delta\pi)$ from
the Luttinger sum rule. The charge phase with gapless  excitation
 forms a Tomonaga-Luttinger ground state. According to numerical works
 \cite{Tsunetsugu}\cite{troyer},
  holes have lower energies when they form pairs across the rungs of the ladders.
 Thus, low-energy excitations include hole-hole pairs and
 hole-spin  bound states (with positive charge)
 propagating along the ladder with a kinetic energy of about $4t_{\parallel}^2/J_{\perp}$.
 These excitations as well as the fluctuations
 of hole pairs move against the lowest-energy background of charge-density wave(CDW).  
 The holon speed is  $v_{h}=2t_{\parallel}sin\delta\pi$.
 
 \section{ correlations}
 
  The asymptotic electron Green's function  can be calculated from perturbation theory
excluding the vertex corrections\cite{wen}\cite{Finkelstein} and is given by
\begin{equation}
G_{e}\approx  {e^{i2k_{F}x}\over (x^2-v_{s}^{2}t^{2})^{\gamma}(x-v_{h}t)^{\eta}
(x+v_{h}t)^{\eta}}\end{equation}
where $\gamma=({\sqrt {1+W^2}-1)/2}$ 
 with a spinon velocity  $v_{s}= v_{F}\equiv 1$ and $\eta=1/2$. The propagator $G_{e}$ can be determined
 by convoluting the holon propagator and the spinon 
 propagator and has no simple poles for low-energy states. Thus, the Fermi
surfaces are considered separately for the holon and for the spinon. The size of the holon Fermi surface
 is found from $\sum_{k,\omega} G_{h}(k, \omega)=\delta$ ($\delta$ being the doping concentration). 
 The ground state of the half-filled insulator is a disordered (or dimerized)
spin liquid made of spin singlets. These singlets which
  form across the rungs  become spin triplets as they are excited.
 Also, the triplets can propagate  along the ladder as well as
  along the chain. The propagation ${\it across}$ the rungs and ladders are ``incoherent"
  in the sense that
 there is no stable eigenstate for the electron. The quantum numbers, such as spin,
 charge, and momentum, are not the same throughout the propagation from one leg to
 another\cite{clarke}. As separated phase fields, the charge and the spin degrees of freedom propagate with different
  speeds. At small lengths, the phase separation may manifest itself spatially 
   in real space, for example as striped phases, as discussed by some authors\cite{Emery}.
    To see whether there is any strong-coupled fixed point, consider turning off the
   interchain hopping $t_{\perp}$. 
   The renormalization group(RG) dimension of the wavefunction comes from rescaling
   the  variables $(x,y,t)\rightarrow (\alpha x, y, \alpha t)$. From the propagator, we have
   $\psi\rightarrow \alpha^{-{1\over 2}-\eta}$. So,  the Hamiltonian
   $H_{\perp}$ has the RG dimension of ${1\over 2} + \eta$. 
   The interchain hopping is irrelevant
   and the Luttinger liquid becomes a fixed point, if and only if $\eta > {1\over 2}$. 
  This corresponds to the case of  interchain interactions being much weaker than
 intrachain ones ($t_\perp\gg t_\parallel$).
 However, in strongly coupled ladders, $J_\perp/J_\parallel$ remains 
 sufficiently large to develop gaps\cite{Khveshchenko}\cite{sachdev} and RG flow is away from the unstable Luttinger liquid fixed point\cite{fabrizio}. 
 Given that $t_{\perp}$ is greater than the spin gap, there is no strong-coupled fixed point, for 
   even-legged ladders($\eta \le {1\over 2})$. 
  The equal-time spin-spin correlation is a power-law form
\begin{equation}
<S(r)S(0)> \approx x^{-\sqrt {1+W^2}}\hspace{0.1cm}cos\hspace{0.1cm}2k_{F}x,
\end{equation}
which shows the spinons being a non-Fermi liquid.
From  second-order perturbation theory\cite{reigrotzki}, 
 the spin gap size is 
$ J_{\perp} + J_{\parallel}cos\hspace{0.1cm}k_{\parallel} + 
{1\over 4} J^2_{\parallel}/J_{\perp}(3-cos\hspace{0.1cm}2k_{\parallel})$
with $k_{\parallel}=\pi$.

\section{ deconfinement}

     In the 2D AF Heisenberg model or $CP^{1}$ model, deconfinement of the electron (to a
 holon and spinon) occurs at $T=0$\cite{ichinose}. 
 As for spin-ladder systems, the ``electron'' phase is realized by having the spin and
 the charge phases confined  at finite $T$ above
 the transition temperature $T_{CD}$. Since the confinement-deconfinement(CD) transition
 occurs in the absence of ordering, it is like a Kosterlitz-Thouless 
 transition occuring in (2+1) dimensional system.
From the partition function, $T_{CD}$ is found
\begin{equation}
{\delta_{\parallel} t_{\parallel}-J_{\parallel}\over 2k[\delta_\perp t_{\perp}-J_{\perp}]}< T_{CD}
< {\delta_{\parallel} t_{\parallel}-J_{\parallel}\over k[\delta_\perp t_{\perp}-J_{\perp}]}
\end{equation}
where $k$ is Boltzmann 
constant. Hence, for $T > T_{CD}$, the electron propagator has the 
product form  $G_{s}(r,t)*G_{h}(-r,-t)$. Below $T_{CD}$, the spin phase and the charge
phase are separated
 due to the screening of the confining potential between holons and spinons.
 Spin-charge separation is 
 a well-accepted fact for 1D systems which are Luttinger liquids. 
 Spin ladders  make transition, at some doping rates  $\delta$,
 to a Luttinger liquid phase,
 where both spin and charge excitations become gapless.

  In summary,  we map the spinons to
  solitons which are  either magnons (excitons when bound) or  excited phasons.
  From the soliton gap, we have the spin-gapped mode while the charge part is
  gapless Luttinger liquid. As the doping increases the ladders can develop
  superconducting phase or lose the spin gap beyond certain hole concentration.
   Also, depending on  the ratio of
  interchain couplings to intraleg interactions, the spin ladders can become
  gapless Luttinger liquid. Our phase diagram  is consistent with the features found in MF studies
  or weak-coupling approaches, such as presence of
  gapless mode and superconducting  phase from even-ladders. 
  The comprehensive diagram shows  possible
  transitions  as an explicit function of coupling ratio and doping.
The  electron propagator  and the spin-spin correlation 
 obey power-laws with the exponents depending on the interaction strengths and the doping rate.
 This shows the correlation of topological defects and phase excitations.
    Constructive interferences of phase excitations (solitons 
  and magnons) may  lead to the formation of 
  intermediate-range or long range order, which is our forthcoming study.  The 
   electron is seen as a confined phase of the gauge flux associated with each point.
   Below the estimated transition temperature, the spin phase and the charge phase are
  separated. Henceforth, even-ladders show the crossover from gapped 2D-like states
  to gapless 1D-nature, depending on the interleg couplings and/or the doping rate.

\acknowledgments

 Y. Bai is grateful to 
 Dr. Chul Koo Kim for suggesting the problem and to M. Sigrist for invaluable discussions. This work is supported by the
 Creative Research Initiative program of Ministry of Science and Technology, Korea.\\

\appendix 
\section{renormalization group equations}
 
 Following Fabrizio\cite{fabrizio}, Kishine\cite{kishine} and Bourbonnais\cite{bourbonnais}, we investigate 
  the aymptotic behaviours in/around the gapped phase by RG flow of the couplings. 
 The partition function $Z=\int D e^S$ where the action consists of one-particle hopping process $S_{(1)}$ and two-particle scattering $S_{(2)}$ in the ladder,  $S= S_{(1)} + S_{(2)}$. The dispersion is linearized with respect to the bonding(B) and anti-bonding(A) band which lie above(inner) the bonding band(LB, LA, RA, RB), where $L$, $R$ denote the right-moving, left-moving fields. Each band ranges by the bandwidth cutoff $K_0$.  The interladder repulsion generates 
  scattering vertices of Fig. 3.
 The action for the one-particle process is given
  \begin{equation}
 S_{(1)}=\sum_{K,\sigma}\sum_{\eta=A,B}[G^{-1}_{L\eta}(K_{\parallel})
 L^+_{\eta\sigma}(K)L_{\eta\sigma}(K)+ G^{-1}_{R\eta}(K_{\parallel})
 R^+_{\eta\sigma}(K)R_{\eta\sigma}(K)
\end{equation}
 where $K=(k_\parallel, k_\perp, i\epsilon)$, $K_\parallel=(k_\parallel,  i\epsilon)$ with the fermion frequency
  $\epsilon_n=(2n+1)\pi\beta$, and the Green's functions are 
  $G_{L\eta}=[i\epsilon_\eta+v_F(k_\parallel+ k_{F\eta})]^{-1}$,
   $G_{R\eta}=[i\epsilon_\eta-v_F(k_\parallel- k_{F\eta})]^{-1}$,
  with the Fermi momenta of the anti-bonding band 
  $k_{FA}=k_F-t_\perp/v_F$ and of the bonding band   
  $k_{FB}=k_F + t_\perp/v_F$. The action for the two-particle processes is given 
  \begin{eqnarray}
  S_{(2)} & = & {2\pi v_F\over \beta}\sum_{\eta,K,\sigma}[g_{0} (\sigma_1 \sigma_2\sigma_3\sigma_4)L^+_{\eta\sigma_1}(K_1)
   R^+_{\eta\sigma_2}(K_2)R_{\eta\sigma_3}(K_3)L_{\eta\sigma_4}(K_4)\delta(K_1+K_2-K_3-K_4)\nonumber\\
   & + & g_{f}(\sigma_1 \sigma_2\sigma_3\sigma_4)L^+_{\eta\sigma_1}(K_1)
   R^+_{\eta\prime\sigma_2}(K_2)R_{\eta\prime\sigma_3}(K_3)L_{\eta\sigma_4}(K_4)\delta(K_1+K_2-K_3-K_4)
   \nonumber\\
     & + & g_{b}(\sigma_1 \sigma_2\sigma_3\sigma_4)L^+_{\eta\sigma_1}(K_1)
   R^+_{\eta\prime\sigma_2}(K_2)L_{\eta\sigma_3}(K_3)R_{\eta\prime\sigma_4}(K_4)\delta(K_1+K_2-K_3-K_4)
   \nonumber\\
     & + & g_{t}(\sigma_1 \sigma_2\sigma_3\sigma_4)L^+_{\eta\sigma_1}(K_1)
   R^+_{\eta\sigma_2}(K_2)R_{\eta\prime\sigma_3}(K_3)L_{\eta\prime\sigma_4}(K_4)\delta(K_1+K_2-K_3-K_4)]
\end{eqnarray}
where the dimensionless couplings $g(\sigma_1 \sigma_2\sigma_3\sigma_4)=
g^{(b)}\delta(\sigma_1\sigma_3)\delta(\sigma_2\sigma_4)-
g^{(f)}\delta(\sigma_1\sigma_3)\delta(\sigma_2\sigma_4)$ and $g_0$ is for the intraband process, $g_f$ for interband forward scattering, $g_b$ for interband backward scattering, $g_t$ for interband tunneling. 
We scale the two-spin processes to remove the high-energy components, by parametrizing the bandwidth cutoff $K(l)=K_0 e^{-l}$ with length $l$.  The scaling equations for the scattering vertices are from
 ${d\ln{g_\alpha}^i(l)\over dl}=z_{\alpha}^i(l)-2z(l)$ where $i=b$ for backward scattering and $i=f$ for forward scattering. The vertex correction diagrams to the second order of the couplings($Z_{\alpha}^i(l)$) are shown in Fig. 4. 
 The linearized bands consist of the slow modes(s) (from the central portions of the branches) and the fast modes(f) (from the outer edges). The action is decomposed $S_{(1),s}+S_{(2),s}+S_{(1),f}+S_{(2),f}$. Expanding the fast components $\exp[S_{(1),f}+S_{(2),f}]$ with respect to $S_{(2),f}$ and integrating out with the fast modes over $dK(l)={1\over 2}K(l) dl$, 
  the partition function becomes 
  $Z=\int D_s{\exp[S_{(1),s}+S_{(2),s}+\sum^\infty_{l=0}{1\over l!}<<S_{(2),f}^l>>_c}]$ where $<< >>_c$ denotes the connected diagram.
   The renormalized propagator and hopping amplitudes give renormalized action.
  \begin{equation}
  \widetilde{G}^{-1}_{\alpha\eta}(K_\parallel)=[1+z(l)dl +\cal{O}\rm{(dl^2)}]
  G^{-1}_{\alpha\eta}(K)
\end{equation}
\begin{equation}
\widetilde{g}^{i}_{\alpha}=[1+z^{i}_{\alpha}(l)dl +\cal{O}\rm{(dl^2)}]
  g^{i}_{\alpha}
\end{equation}
\begin{equation}
\widetilde{t_\perp}=[1+\cal{O}\rm{(dl^2)}]]t_\perp
\end{equation}
where $\alpha$ stand for (o, f, b, t) and $i$ for backward(1) and forward(2) scatterings. The Feynman diagrams contributing to the order $dl$ are evaluated. Those in Fig. 4 give $z^{i}_{\alpha}(l)$, while the self-energy diagrams give $z(l)$ of the scaling equations (Fig. 5).
\begin{eqnarray}
z(l)&=& g^{(1)}_0(l)^2 + g^{(2)}_0(l)^2 + g^{(1)}_f(l)^2 + g^{(2)}_f(l)^2 +  g^{(1)}_b(l)^2 + g^{(2)}_b(l)^2 +  g^{(1)}_t(l)^2 \nonumber\\
&+& g^{(2)}_t(l)^2 -g^{(1})_0(l)g^{(2)}_0(l)- g^{(1)}_f(l)g^{(2)}_f(l)- g^{(1)}_b(l)g^{(2)}_b(l)
  - g^{(1)}_t(l)g^{(2)}_t(l)
\end{eqnarray}
The scaling equations follow, provided small $K_0 < \Delta k_F$ ;
\begin{eqnarray}
{d\over dl}g^{(1)}_0(l)&=&-2(g^{(1)}_0(l)^2 +g^{(1)}_t(l)g^{(2)}_t(l)) + 
2(g^{(1)}_f(l)g^{(2)}_t(l)^2-g^{(1)}_f(l)g^{(1)}_t(l)g^{(2)}_t(l)) 
\nonumber\\
&-&
2g^{(1)}_0(l)(g^{(1)}_0(l)^2+ g^{(1)}_f(l)^2 + g^{(1)}_t(l)^2 +g^{(2)}_t(l)^2- g^{(1)}_t(l)g^{(2)}_t(l))
\end{eqnarray}
\begin{eqnarray}
{d\over dl}g^{(2)}_0(l)&=&-(g^{(1)}_0(l)^2 +g^{(1)}_t(l)^2+g^{(2)}_t(l)^2) + 
2(g^{(2)}_f(l)g^{(2)}_t(l)^2-g^{(1)}_t(l)g^{(2)}_t(l)g^{(2)}_f(l))
\nonumber\\
&+& 2g^{(1)^{2}}_{t}(l)g^{(2)}_{f}(l)-g^{(1)^{3}}_{0}-g^{(1)^{2}}_{f}(l) g^{(1)}_{0}(l)-
g^{(1)^{2}}_{t}(l)g^{(1)}_{f}(l)
\nonumber\\
&-& 2g^{(2)}_0(l)(g^{(1)}_t(l)^2+ g^{(2)}_t(l)^2 - g^{(1)}_t(l)g^{(2)}_t(l))
\end{eqnarray} 
\begin{eqnarray}
{d\over dl}g^{(1)}_f(l)&=&-2(g^{(1)}_f(l)^2 +g^{(1)}_t(l)^2 -g^{(1)}_t(l)g^{(2)}_t(l))+2[g^{(1)}_0(l)g^{(2)}_t(l)^2-g^{(1)}_0(l)g^{(1)}_t(l)g^{(2)}_t(l)]
\nonumber\\
 &-& 2g^{(1)}_f(l)[g^{(1)}_0(l)^2+g^{(1)}_f(l)^2+g^{(1)}_t(l)^2
g^{(2)}_t(l)^2-g^{(1)}_t(l)g^{(2)}_t(l)]
\end{eqnarray} 
\begin{eqnarray}
{d\over dl}g^{(2)}_f(l)&=&-(g^{(1)}_f(l)^2 -g^{(2)}_t(l)^2)+
2(g^{(2)}_0(l) g^{(2)}_t(l)^2-g^{(1)}_t(l)g^{(2)}_t(l)g^{(2)}_0(l))\nonumber\\
&+&2g^{(1)}_t(l)^2g^{(2)}_0(l)
-g^{(1)}_t(l)^2g^{(1)}_0(l)-g^{(1)}_0(l)^{2}g^{(1)}_f(l)-g^{(1)}_f(l)^3\nonumber\\
&-&
2(g^{(2)}_f(l)(g^{(1)}_t(l)^2 +g^{(2)}_t(l)^2-g^{(1)}_t(l)g^{(2)}_t(l))
\end{eqnarray} 
\begin{eqnarray}
{d\over dl}g^{(1)}_b(l)&=&-2(g^{(1)}_b(l)g^{(2)}_f(l)+g^{(1)}_f(l)g^{(2)}_b(l)
 +g^{(1)}_b(l)g^{(1)}_0(l)-g^{(1)}_b(l)g^{(2)}_0(l)\nonumber\\
 &+&g^{(1)}_0(l)g^{(1)}_b(l)-
 g^{(1)}_0(l)g^{(2)}_b(l))-2g^{(1)}_b(l)[(g^{(2)}_b(l)-g^{(2)}_f(l))^2\nonumber\\
&-& g^{(1)}_0(l)g^{(2)}_0(l)+g^{(1)}_0(l)g^{(2)}_f(l)+g^{(1)}_f(l)g^{(2)}_0(l) - g^{(1)}_f(l)g^{(2)}_f(l)+g^{(1)}_0(l)^2+g^{(1)}_f(l)^2]
\nonumber\\
& +& 2g^{(1)}_b(l)(g^{(1)}_t(l)^2-g^{(1)}_t(l)g^{(2)}_t(l)
+g^{(2)}_t(l)^2)
 \end{eqnarray} 
\begin{eqnarray}
{d\over dl}g^{(2)}_b(l)&=&-2(g^{(2)}_b(l)g^{(2)}_f(l)-g^{(2)}_b(l)g^{(2)}_0(l)+g^{(1)}_b(l)g^{(1)}_f(l))\nonumber\\
& -& 2g^{(1)}_b(l)g^{(1)}_0(l)g^{(1)}_f(l)
  -2g^{(2)}_b(l)[(g^{(1)}_0(l)-g^{(1)}_f(l))^2-
 g^{(1)}_0(l)g^{(2)}_0(l)\nonumber\\
& + & g^{(1)}_0(l)g^{(2)}_f(l)+g^{(1)}_f(l)g^{(2)}_0(l)
 - g^{(1)}_f(l)g^{(2)}_f(l)+(g^{(2)}_0(l)-g^{(2)}_f(l))^2]
 \nonumber\\
 &-& 2g^{(2)}_b(l)(g^{(1)}_t(l)^2-g^{(1)}_t(l)g^{(2)}_t(l)
 +g^{(2)}_t(l)^2)
\end{eqnarray} 
\begin{eqnarray}
{d\over dl}g^{(1)}_t(l)&=&-2(2g^{(1)}_t(l)g^(1)_f(l)-
g^{(1)}_t(l)g^{(2)}_f(l)
 -g^{(1)}_f(l)g^{(2)}_t(l)+g^{(1)}_0(l)g^{(2)}_t(l)\nonumber\\
& +& g^{(1)}_t(l)g^{(2)}_0(l))
+2(2g^{(1)}_t(l)g^{(2)}_0(l)g^{(2)}_f(l)-g^{(1)}_0(l)g^{(1)}_t(l)g^{(2)}_f(l)\nonumber\\
&-& g^{(1)}_f(l)g^{(1)}_t(l)g^{(2)}_0(l))-2g^{(1)}_t(l)(g^{(1)}_0(l)^2
 + g^{(2)}_0(l)^2 +g^{(1)}_f(l)^2+g^{(2)}_f(l)^2\nonumber\\
& +& g^{(1)}_t(l)^2+g^{(2)}_t(l)^2-
  g^{(1)}_0(l)g^{(2)}_0(l)- g^{(1)}_f(l)g^{(2)}_f(l)-g^{(1)}_t(l)g^{(2)}_t(l))
  \end{eqnarray} 
 \begin{eqnarray}
 {d\over dl}g^{(2)}_t(l)&=&-2(g^{(1)}_0(l)g^{(1)}_t(l)-
 g^{(2)}_0(l)g^{(2)}_t(l) -g^{(2)}_f(l)g^{(2)}_t(l))\nonumber\\
 &+& 2(2g^{(2)}_0(l)g^{(2)}_f(l)g^{(2)}_t(l)-
 g^{(1)}_f(l)g^{(2)}_0(l)g^{(2)}_t(l)-
  g^{(1)}_0(l)g^{(2)}_f(l)g^{(2)}_t(l))\nonumber\\
 & +& 2g^{(1)}_0(l)g^{(1)}_f(l)g^{(2)}_t(l)-
  g^{(1)}_t(l)g^{(1)}_f(l)g^{(1)}_0(l))
 -2g^{(2)}_t(l)(g^{(1)}_0(l)^2+g^{(2)}_0(l)^2+g^{(1)}_f(l)^2
 \nonumber\\
&+& g^{(2)}_f(l)^2+ g^{(1)}_t(l)^2+g^{(2)}_t(l)^2
   -g^{(1)}_0(l)g^{(2)}_0(l)- g^{(1)}_f(l)g^{(2)}_f(l)-g^{(1)}_t(l)g^{(2)}_t(l))
   \end{eqnarray} 
 From the initial condition $g_{\alpha}^{(i)}(0)=t^2/{J\pi v_F}\equiv I
 \ge 0$, the scaling equations give the strong-coupling fixed point ;
 $g_0^{(1)}=-1$,  $g_f^{(1)}=0$,  $g_t^{(1)}=1$,
   $g_0^{(2)}=1/4(2I-3)$,  $g_f^{(2)}=1/4(2I+1)$, 
   $g_f^{(2)}=1$. However, the couplings actually do not flow to this fixed point, as far as the interchain hopping is greater than the gap size, since  interchain pair fluctuations are dominant excitations\cite{fabrizio}. The scale invariant coupling for the charge part is
 $g_{\rho,\pm}=(g^{(1)}_0-2g^{(2)}_0)\pm (g^{(1)}_f-2g^{(2)}_f)$.
  The invariant coupling for the spin part 
  $g_{\sigma,\pm}=g^{(1)}_0\pm g^{(1)}_f$ where the $+$ sign is for the total mode and $-$ for the relative mode.  From above g's, the spin stiffness is determined $K_\mu=\sqrt{{1+g_\mu}\over{1-g_\mu}}$. 
  So, $K_{\sigma,+}=0$ means that the system developed the spin gap phase, while the charge part remains gapless from $K_{\rho,\pm}=0$
  (Luther-Emery phase)\cite{lin}.  In our equation (8), the sine-Gordon mass term identifies $W\sim (Jv_F/t^2)^4{g_1\over \sqrt{v_F^2-g_1^2}}$ where $g_1$ respresents some strong coupling value resulted from the scaling equations.

 \begin{figure}
 \begin{center}
 \epsfig{file=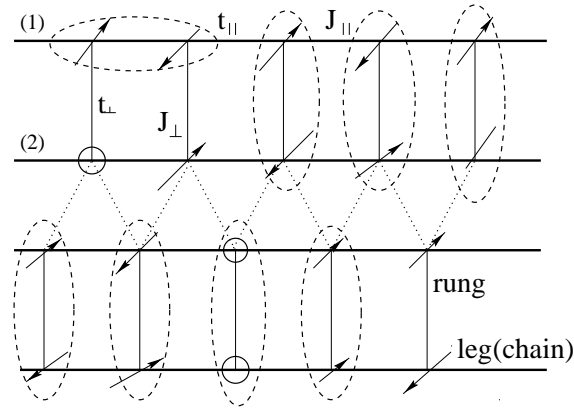,clip=,width=0.3\textwidth,angle=270}
\end{center}
 \caption{spin ladder example : a two-leg ladder system. The dashed 
 circles are for spin singlets (triplets, if excited) and for 
  holon pairs after doped.}
 \label{fig1}
\end{figure}

\begin{figure}
\begin{center}
\epsfig{file=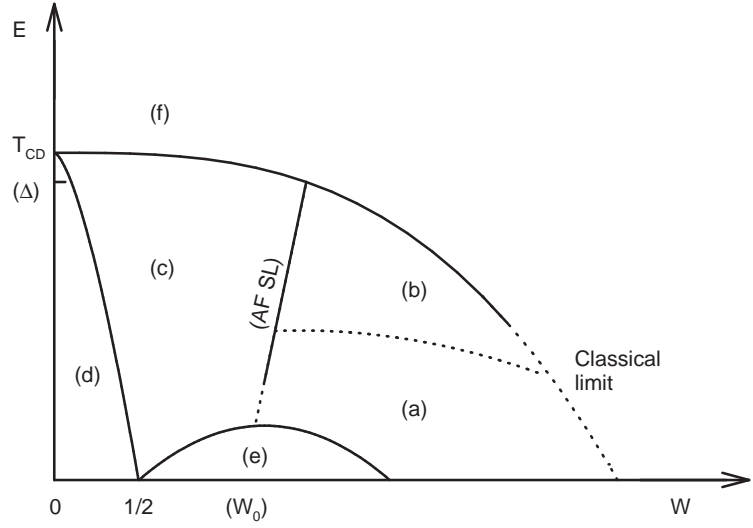,clip=,width=0.7\textwidth}
\end{center}
\caption{phase diagram : the energy scale vs. the ratio 
$W=[\delta_{\perp} t_{\perp}-J_{\perp}]/[\delta_{\parallel} t_{\parallel}-J_{\parallel}]$\\
``AF SL'' stands for antiferromagnetic spin liquid which occurs at
 $W=W_0=J_{\perp}/J_{\parallel}$ with $\delta=0$. The spin gapsize $\Delta$
 is marked for the point where gapped spinon phases branch from AF SL state.
(a)gapped magnons; excitonic bound states of spinons,Luther-Emery liquid(LE)
 with gapless charge excitation (b) LE with gapful spinons and gapless holons, CDW 
(c) unstable solitons, gauge bosons, SDW (d) Luttinger liquid (paramagnetic); gapless
excitations of both-charge and spin (e) superconducting (with d-wave symmetry)
(f) electron phase:  spinon and holon confined}
\label{fig2}
\end{figure}
\begin{figure}
\begin{center}
\epsfig{file=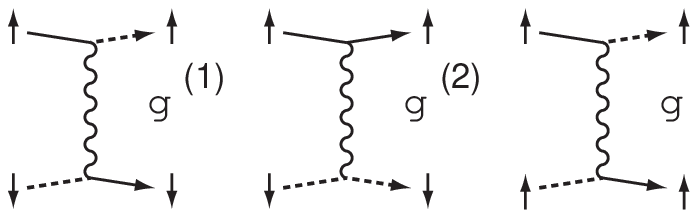,clip=,width=0.7\textwidth}
\end{center}
\caption{interaction vertices. The solid lines are for the right-going fermions and the dashed lines for the left-going ones. $g^{(1)}$ and $g^{(2)}$ denote  backward and forward scattering couplings, respectively.} 
\label{fig3}
\end{figure}

\begin{figure}
\begin{center}
\epsfig{file=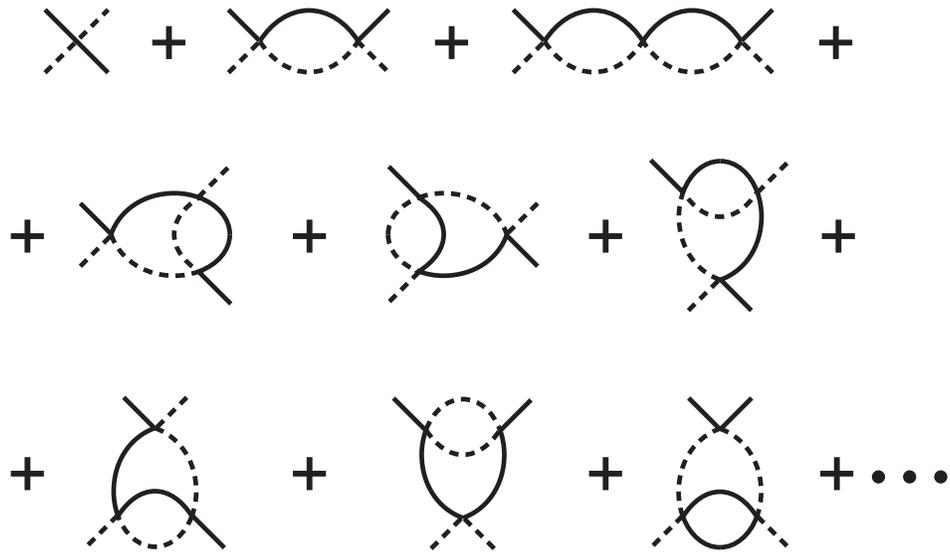,clip=,width=0.7\textwidth}
\end{center}
\caption{vertex diagrams to the next-to-the-leading order.}
\label{fig4}
\end{figure}

\end{document}